\documentclass[aps,prl,twocolumn,amsmath,amssymbs,superscriptaddress,longbibliography]{revtex4-1}
\usepackage{epsfig}
\usepackage{wrapfig}
\usepackage{bbm}
\usepackage[usenames]{color}
\usepackage{array}
\usepackage{times}

\usepackage{graphicx}
\usepackage{float}
\usepackage{multirow}
\usepackage{natbib,twoopt}
\usepackage[breaklinks=true]{hyperref}
\usepackage{xcolor}
\usepackage{amsmath,amsfonts,amssymb}
\usepackage{graphicx}
\usepackage{hyperref}
\usepackage{color}
\usepackage{stackengine}
\usepackage{subfigure}
\usepackage{verbatim}
\usepackage{stmaryrd} 


\usepackage{soul}

\newcommand{\RNum}[1]{\uppercase\expandafter{\romannumeral #1\relax}}

\newcommand{\balancecolsandclearpage}{%
	\close@column@grid
	\cleardoublepage
	\twocolumngrid
}

\begin{document}

\title{Crafting crystalline topological insulators via accidental mode degeneracies}

\author{Konstantin Rodionenko}
\affiliation{School of Physics and Engineering, ITMO University, Saint  Petersburg 197101, Russia}

\author{Maxim Mazanov}
\affiliation{School of Physics and Engineering, ITMO University, Saint  Petersburg 197101, Russia}

\author{Maxim A. Gorlach}
\email{m.gorlach@metalab.ifmo.ru}
\affiliation{School of Physics and Engineering, ITMO University, Saint  Petersburg 197101, Russia}

\begin{abstract}
Crystalline topological insulators have recently become a powerful platform for realizing photonic topological states from microwaves to the visible. Appropriate geometric symmetries of the lattice are at the core of their functionality. Here we put forward an alternative approach to craft those systems by designing the internal symmetries of the Hamiltonian via accidental mode degeneracies. We illustrate our approach constructing an analog of breathing honeycomb lattice using simpler lattice geometry and six times less meta-atoms, reveal edge and corner states and calculate the relevant topological invariants.
\end{abstract}

\maketitle

\textit{Introduction.}~-- Since their inception, photonic topological insulators have provided a viable platform to realize a plethora of topological phenomena~\cite{Lu2014Nov,Ozawa2019Mar,Khanikaev2017,Xie_HOTI_review_2021}. Photonic platform allows one to explore vast frequency range from microwaves to the visible utilizing such physical systems as waveguide arrays~\cite{Rechtsman2013,BlancoRedondo2016,Noh2018,corner2019,Schulz2022Nov,Kang2023Mar}, photonic crystals and metamaterials~\cite{Khanikaev2013Mar,Cheng2016,Slobozhanyuk2017Feb,Yang2019,Bisharat2019}, polaritonic micropillar lattices~\cite{Karzig2015,Bloch_PRL_2017,Klembt2018} and coupled ring resonator arrays~\cite{Hafezi2011,Hafezi2013,Mittal_2019}. To date, most of fabrication-friendly topological systems scalable towards optical wavelengths~\cite{Barik2018Feb,Noh2018,Smirnova2019,Para2020,Vakulenko2021} rely on the specific crystalline lattice geometries  and therefore are called crystalline topological insulators~\cite{Quantization_2019}. 

A profound example of such kind is provided by the well-celebrated breathing honeycomb lattice~\cite{Wu_Hu}. As first predicted in Ref.~\cite{Wu_Hu}, $C_6$-preserving deformation of the honeycomb lattice opens a complete photonic band gap in the dispersion. Depending on the geometric parameters of the lattice, the gap may become topological hosting edge- and corner-localized disorder-robust modes. Remarkable simplicity and scalability of this platform have immediately captured wide interest resulting in multiple experimental realizations from microwave~\cite{Yves2017,Li2018,Yang2018,Xie2020Jul} to optical~\cite{Barik2018Feb,Noh2018,Gorlach2018,Smirnova2019,Para2020,Liu2020,Li2021} domain. 

While lattice geometry in this approach is carefully tailored, the meta-atoms themselves are assumed single-mode, which restricts effective spin degrees of freedom and thus limits  available topological phenomena. Much larger flexibility in constructing and manipulating effective spin degrees of freedom is achieved using multimode meta-atoms with engineered mode degeneracies. 

The simplest way to achieve such degeneracy is to exploit light polarization, which is extensively harnessed in topological photonics~\cite{Khanikaev2013Mar,Slobozhanyuk2017Feb,Bloch2017Oct,Bobylev2022}. However, this also limits pseudo-spin degrees of freedom because of the two available polarizations of light.


Less intuitive but promising approach is to tailor the degeneracy of the different {\it orbital} modes of the meta-atom. Since symmetry of the degenerate modes can be chosen at will, this unlocks a vast realm of topological models. 

Recently, this strategy has been explored for one-dimensional systems. Using the interplay between symmetric ($s$) and antisymmetric ($p$) modes of a single meta-atom, several groups predicted~\cite{Aravena_PRA_2020,SavelevGorlach_PRB,Jiang2023,Gorbach:23} and observed experimentally~\cite{Mikhin2023,Mazanov2023Oct,Liu2023Oct} topological edge states in an equidistant array arising due to the inter-orbital coupling~\cite{Silva_PRL_2021}.

Even richer physics is expected in two-dimensional arrays of multimode meta-atoms whose physics remains barely charted with only few first studies currently available~\cite{Mazanov2022May,Mazanov2022,Lu2020May}.

\begin{figure}[b!]
    \centering
    \includegraphics[width = 0.75\linewidth]{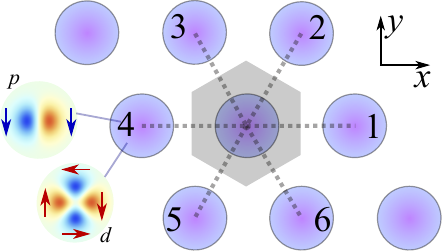}
    \caption{Design of the triangular periodic lattice with nearly degenerate $p_\pm$ and $d_\pm$ orbital modes. $C_6$-symmetric unit cell is highlighted in gray. Color encodes the phase of the eigenmode electric field.
    }
	\label{Fig1}
\end{figure}

In this Letter, we make a conceptual step in this direction showing that a simple triangular lattice of two-mode waveguides with the degenerate $p$ and $d$ orbital modes [Fig.~\ref{Fig1}] features the same topological properties as a crystalline topological insulator based on the breathing honeycomb lattice~\cite{Wu_Hu}. In addition to reproducing topological edge and corner states  with $6$ times less meta-atoms, our model incorporates an additional tuning mechanism via on-site meta-atom properties.

Below, we develop an analytical tight-binding description, calculate the relevant topological invariants and corroborate this description by the full-wave numerical simulations.

\textit{Model and band structure.}~-- 
We consider a triangular periodic lattice consisting of evanescently coupled waveguides supporting two accidentally degenerate pairs of dipolar $(p_x,p_y)$ and quadrupolar $(d_{xy},d_{x^2-y^2})$ modes. The rest of the waveguide modes are neglected assuming that the detuning of their propagation constants from those of $p$ and $d$ orbitals is much larger than the $p$-$d$ inter-orbital coupling.

The tight-binding Hamiltonian $\hat{H}$ of this model can be constructed by calculating the overlap integrals of the respective modes~\cite{SavelevGorlach_PRB}. Using the basis of circularly polarized modes $p_{\pm}=p_x\pm ip_y$, $d_{\pm}=d_{xy}\pm id_{x^2-y^2}$ ordered as $(p_+,p_-,d_+,d_-)^T$ and possessing azimuthally-invariant intensity profiles, we recover~\cite{Supplement}
%

\begin{figure*}[htb]
    \centering
    \includegraphics[width=0.85\textwidth]{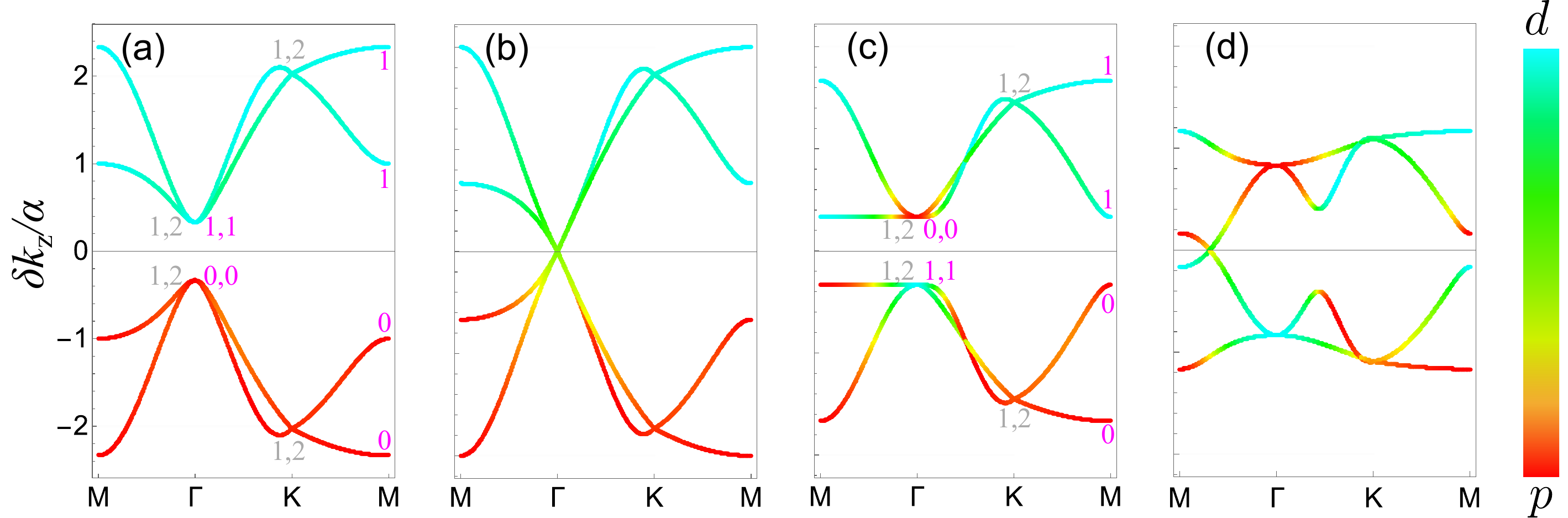}
    \caption{Calculated spectra of propagation constants for (a) $\delta\beta=-8\alpha$, (b) $\delta\beta=-6\alpha$, (c) $\delta\beta=-4\alpha$, (d) $\delta\beta=-\alpha$. Blue and red colors indicate $d$ and $p$ modes, respectively. 
    The indicators for symmetry operators $C_2$ and $C_3$ (numbers $p$ from symmetry operator eigenvalues $\Pi^{(2,3)} = e^{2 \pi i (p-1) / N}$ for $N=2,3$~\cite{Quantization_2019}) at $M$, $K$ and $\Gamma$ points are shown near high symmetry points at panels (a,c). 
    }
    \label{Fig2}
\end{figure*}

\begin{equation}
    \label{Ham}
    \hat{H} = \sum_{j=0}^{6} \hat{\varkappa_j} \, \exp\left(i k_x \cos\frac{2\pi j}{3}  +  i k_y \sin\frac{2\pi j}{3}\right) + \delta\hat{\beta}\:.
\end{equation}
Here $\delta\hat{\beta} = \delta\beta \cdot \hat{\sigma}_z \otimes \hat{I}$ (with $\hat{\sigma}_z$ being the third Pauli matrix) captures the difference between the propagation constants of $p$ and $d$ modes, respectively. $\hat{\varkappa_i}$ are $4 \times 4$ matrices describing the coupling between the orbital modes of the neighboring waveguides where index $i$ labels the direction as depicted  in~Fig.~\ref{Fig1}. These coupling matrices have the following structure~\cite{Supplement} 
\begin{equation}
\hat{\kappa} (\gamma)
=
\left(
\begin{array}{cccc}
 -a & b \gamma^{-1} & c \gamma^2 & d \\
 b \gamma & -a & d & c \gamma^{-2} \\
 c \gamma & -d & e & f \gamma^{-2} \\
 -d & c \gamma^{-1} & f \gamma^{2} & e \\
\end{array}
\right), 
\end{equation}
with $
\hat{\kappa}_1 = \hat{\kappa}^T(e^{i \pi}), 
\hat{\kappa}_2 = \hat{\kappa}(e^{i \pi / 3}), 
\hat{\kappa}_3 = \hat{\kappa}^T_2, 
\hat{\kappa}_4 = \hat{\kappa}_1^\dag, 
\hat{\kappa}_5 = \hat{\kappa}_2^\dag,
\hat{\kappa}_6 = \hat{\kappa}_3^\dag
$. 
%
The coupling matrices $\hat{\kappa_i}$ above are not fully independent and are linked to each other due to $C_6$ lattice symmetry. As a result, all possible couplings in the lattice are captured by $6$ independent real-valued and positive coupling parameters $a$-$f$~\cite{Supplement}, which provides a lot of freedom in constructing topological phases.

Here, for the sake of clarity, we explore the situation when all six coupling parameters are equal to each other, i.e. $a=b=c=d=f=e=\alpha$. Our simulations~\cite{Supplement} suggest that such scenario is readily attainable in experiments, while different choice of $a-e$ parameters yields a similar physics. In such setting, $\alpha$ quantifies the strength of the inter-orbital coupling, while $\delta\beta$ measures the detuning between the propagation constants of $p$ and $d$ modes, respectively. Depending on the ratio between those parameters, we recover different band structures shown in Fig.~\ref{Fig2}.

Our calculations suggest that only negative $\delta\beta$ lead to the nontrivial physics. If the detuning between the modes $\delta\beta$ significantly exceeds the inter-orbital coupling $\alpha$, the bands produced by $p$ and $d$ modes are detached from each other. The bandgap in the spectrum is largely defined by the $\delta\beta$ value and has trivial origin [Fig.~\ref{Fig2}(a)]. However, with the decrease of $\delta\beta$ magnitude while keeping its negative sign the bandgap narrows eventually closing at $\Gamma$ point for $\delta\beta=-6\alpha$ [Fig.~\ref{Fig2}(b)]. In the vicinity of $\Gamma$ point, the spectrum possesses double-Dirac cone structure which aligns with the symmetry arguments~\cite{Sakoda_2012_1,Sakoda_2012_2,Sakoda_2012_3}.

Further decrease of the $|\delta\beta|$ causes the bandgap to reopen [Fig.~\ref{Fig2}(c)]. However, contrary to the topologically trivial scenario, $p$ and $d$ modes now experience a significant mixing. Interestingly, further decrease of $|\delta\beta|$ does not improve the width of the bandgap, but rather closes it again for $\delta\beta=-2\alpha$. This time, the closing of the bandgap occurs between $\Gamma$ and M points as depicted in Fig.~\ref{Fig2}(d). The observed behavior of the spectrum strongly resembles that of the canonical Wu-Hu model~\cite{Wu_Hu} indicating the onset of nontrivial topology.


\textit{Topological properties.}~-- To further examine the topological properties of our model, we calculate the topological invariants defined in terms of bulk eigenstates. To that end, we examine the symmetry of the eigenvectors in a set of high-symmetry points in the Brillouin zone~\cite{Quantization_2019}.

The symmetry group of the entire model is defined by the combination of the lattice symmetry and the symmetry of the orbitals involved. Note that the orbitals $p_{\pm}$ and $d_{\pm}$ possess full rotational symmetry, since their arbitrary rotation results only in an additional phase. Therefore, the entire Hamiltonian possesses $C_6$ symmetry  defined by the point symmetry group of the lattice, and the topological indices need to be computed for this case~\cite{Quantization_2019}.

The indicators for symmetry operators $C_2$ and $C_3$ at $M$, $K$ and $\Gamma$ points are shown in Fig.~\ref{Fig2}(a,c), yielding a nontrivial index $\chi^{(6)}= (2,0)$ in the topological phase [Fig.~\ref{Fig2}(c)] for the range of detunings $-6\alpha < \delta \beta < -2\alpha $ with nontrivial bandgaps. The obtained symmetry indicators are the same as in the breathing honeycomb design~\cite{Wu_Hu}. 

\begin{figure*}[tbp]
    \centering
    \includegraphics[width=0.99\textwidth]{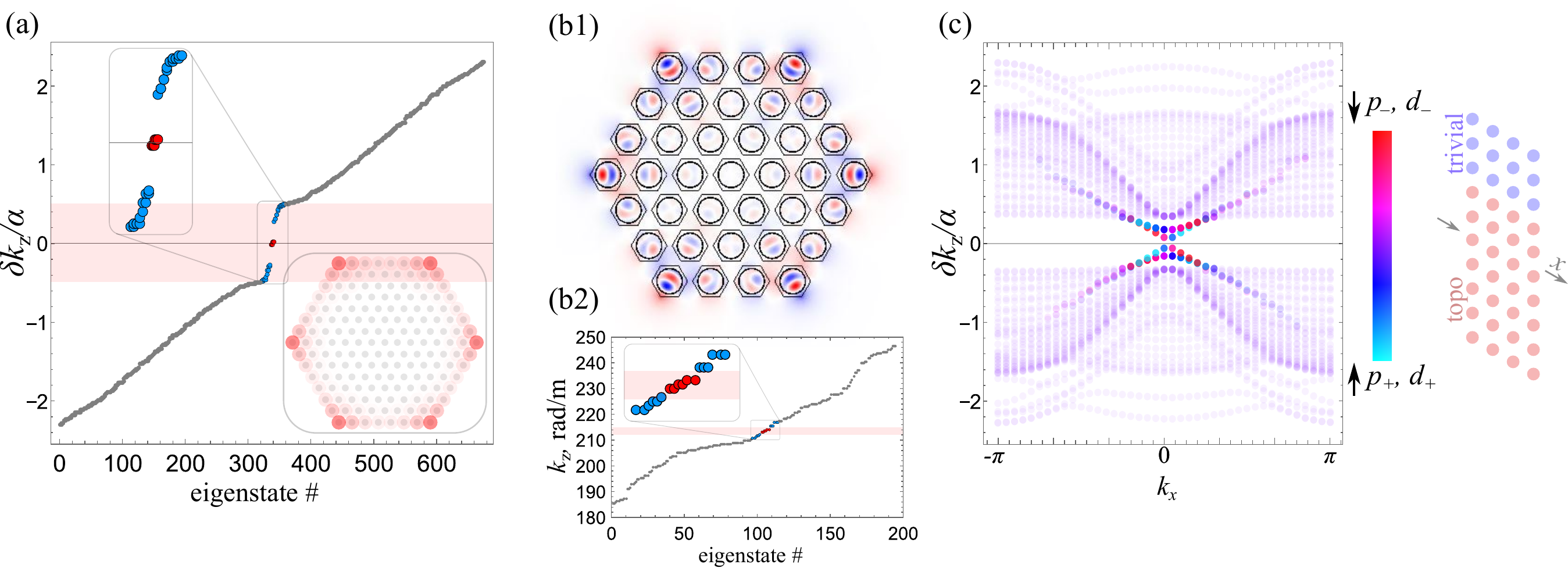}
    \caption{(a-b) Calculated spectra for a finite hexagonal  meta-structure. The topological gap is highlighted by light red. (a) Tight-binding calculation for $\delta \beta = -4\alpha$, inset shows the profile of a corner mode. (b) Full wave numerical simulation of the modes for the array of microwave waveguides~\cite{Supplement} and associated profile ($E_z$ component) of the corner mode. (c) Calculated spectrum in the ribbon geometry. Color marks the pseudospins of the modes: ``pseudospin-down'' $(p_-, d_-)$ and ``pseudospin-up'' $(p_+, d_+)$. Inset shows the ribbon geometry. 
    }
    \label{Fig3}
\end{figure*} 

To further highlight the parallel between our system and the well-celebrated breathing honeycomb geometry and to reveal the onset of photonic quantum Hall physics, we expand the Hamiltonian Eq.~\eqref{Ham} in the vicinity of $\Gamma$ point presenting it in the pseudospin basis $(p_{+}, d_{+}, p_{-}, d_{-})^T$. After discarding cubic and quadratic off-diagonal terms as negligibly small~\cite{Gorlach2018}, we present the Hamiltonian in the block-diagonal form
\begin{equation}\label{eq:HamApr}
\hat{H}=6\alpha\, \begin{pmatrix}
  \mu + k^2/4 &-k_+/2 & 0 & 0 \\
  -k_-/2 & -\mu - k^2/4 & 0 & 0 \\
  0 & 0 & \mu + k^2/4 & k_-/2 \\
  0 & 0 & k_+/2 & -\mu - k^2/4 \\
\end{pmatrix}
, 
\end{equation}
with $k_\pm = k_x \pm i k_y$, $\mu = -\delta\beta/6\alpha - 1$, and blocks having well-defined and opposite Chern numbers $C_\pm = \pm 1/2(\text{sign}(-\delta\beta - 6\alpha)-1)$. These Chern numbers are nonzero provided $\delta\beta > -6 \alpha$, which agrees with the results for the band structure above. This reveals  quantum spin Hall physics and, consequently, predicts the emergence of pseudospin-polarized topological edge states.

Note, however, that the Hamiltonian Eq.~\eqref{eq:HamApr} is approximate and only valid in a small area of reciprocal space around $\Gamma$ point. Therefore, the calculated spin Chern number provides an approximate topological characterization, while rigorous $Z_2$ invariant calculation is obstructed by the lack of band degeneracy at time-reversal-invariant momentum at $M$ point (Fig.~\ref{Fig2})~\cite{Kane_Mele_2005,Kane2005Nov}. Unlike strictly defined set of symmetry indicators $\chi^{(6)}$, the approximate spin-Chern invariant provides a limited protection resulting in the minigaps in the spectrum of the edge states.




{\it Topological edge and corner states.}~-- Nonzero topological invariants predict the emergence of edge and corner states in a finite structure with open boundary conditions. To verify this, we examine the modes of a finite hexagonal sample depicted  schematically in Fig.~\ref{Fig3}. Taking the parameters $\alpha$ and $\delta\beta$ corresponding to topologically nontrivial phase, we compute the spectrum of propagation constants within our tight-binding model [Fig.~\ref{Fig3}(a)]. Some of the computed modes appear in the bulk bandgap shaded in Fig.~\ref{Fig3}(a) by light red. The larger portion of the modes is localized at the edges of the structure. At the same time, we observe six degenerate corner states, one at each obtuse angle of the lattice [Fig.~\ref{Fig3}(a)]. The orbital content of all corner modes is hybrid including all four degenerate modes $p_\pm, d_\pm$. 

To support the tight-binding results, we calculate the spectrum of the finite structure in COMSOL Multiphysics software package exploiting hexagonal waveguides with fine-tuned accidental $p$-$d$ mode degeneracy~\cite{Supplement}. 
The obtained spectrum [Fig.~\ref{Fig3}(b)] features six corner-localized states in the topological bandgap with neighbouring edge states penetrating into the bulk spectrum. One of the corner modes in shown in~Fig.~\ref{Fig3}(b1), with amplitude profile exponentially decaying from the corner and containing a superposition of all four modes, matching the tight-binding predictions. 

In turn, the edge states in our system provide an evidence of the approximate spin Hall physics. To probe it, we simulate the dispersion of the edge states in a ribbon geometry shown in Fig.~\ref{Fig3}(c). As expected, the edge modes with opposite circular polarizations traverse the topological gap featuring an avoided crossing in the vicinity of $\Gamma$ point. Such minigap is attributed to the approximate nature of spin Hall topology in our system.




{\it Discussion and conclusions.}~-- To summarize, our work illustrates the potential of orbital degrees of freedom for tailoring topological phases in two-dimensional lattices. Specifically, creating accidental degeneracy of $p$ and $d$ orbital modes within each individual meta-atom, we recover the physics of the well-celebrated Wu-Hu model. Compared to the original realization based on the breathing honeycomb lattice, our proposal utilizes simpler lattice geometry, requires 6 times less meta-atoms and opens an easy access to continuously tune the topological properties by sweeping the operating wavelength, which governs the detuning between the propagation constants of $p$ and $d$ orbital modes.

We believe that the approach presented here could be generalized towards larger number of degenerate orbitals thus allowing to probe exotic topological models with effective spins $1$ or $3/2$, not available in condensed matter context.

The interplay between complex lattice geometry and on-site orbital degrees of freedom is another promising direction of future research.

{\it Acknowledgments.}~-- 
Theoretical models were supported by Priority 2030 Federal Academic Leadership Program. Numerical simulations were supported by the Russian Science Foundation (Grant No.~23-72-10026).  M.M. and M.A.G. acknowledge partial support from the Foundation for the Advancement of Theoretical Physics and Mathematics ``Basis''.

\bibliography{refs}

\end{document}


\title{Supplemental Materials: \\ Crafting crystalline topological insulators via accidental mode degeneracies}

\author{Konstantin Rodionenko}
\affiliation{School of Physics and Engineering, ITMO University, Saint  Petersburg, Russia}

\author{Maxim Mazanov}
\affiliation{School of Physics and Engineering, ITMO University, Saint  Petersburg, Russia}

\author{Maxim A. Gorlach}
\affiliation{School of Physics and Engineering, ITMO University, Saint  Petersburg, Russia}

\maketitle

\onecolumngrid

\setcounter{equation}{0}
\setcounter{figure}{0}
\setcounter{table}{0}
\setcounter{page}{1}
\setcounter{section}{0}
\makeatletter
\renewcommand{\theequation}{S\arabic{equation}}
\renewcommand{\thefigure}{S\arabic{figure}}
\renewcommand{\bibnumfmt}[1]{[S#1]}
\renewcommand{\citenumfont}[1]{S#1}

\tableofcontents

\section{Supplemental Note 1: Details on coupled-mode equations}

Coupled-mode equations for an array of two-mode waveguides have been derived in Ref.~\cite{SavelevGorlach_PRB} (see Supplementary materials, Section~I of that paper) and generalized to the two-dimensional case in Ref.~\cite{Mazanov2022May}. 
Following those works, we utilize orthogonality of the modes of a single waveguide and the fact that both modes decay exponentially outside of the waveguide. Choosing the basis of circularly polarized $p_\pm$ and $d_\pm$ modes modes, we recover the following system of coupled-mode equations:
\begin{equation}
\frac{d}{d z}\left(\begin{array}{c}
a_{m,n}^{(p_+)} \\
a_{m,n}^{(p_-)} \\ 
a_{m,n}^{(d_+)} \\ 
a_{m,n}^{(d_-)} 
\end{array}\right)=i \sum_{k,l=\pm 1}\left(\begin{array}{cccc}
\kappa_{m,n; m+k, n+l}^{(p_+,p_+)} & \kappa_{m,n; m+k, n+l}^{(p_+,p_-)} & \kappa_{m,n; m+k, n+l}^{(p_+,d_+)} & \kappa_{m,n; m+k, n+l}^{(p_+,d_-)} \\
\kappa_{m,n; m+k, n+l}^{(p_-,p_+)} & \kappa_{m,n; m+k, n+l}^{(p_-,p_-)} & \kappa_{m,n; m+k, n+l}^{(p_-,d_+)} & \kappa_{m,n; m+k, n+l}^{(p_-,d_-)} \\
\kappa_{m,n; m+k, n+l}^{(d_+,p_+)} & \kappa_{m,n; m+k, n+l}^{(d_+,p_-)} & \kappa_{m,n; m+k, n+l}^{(d_+,d_+)} & \kappa_{m,n; m+k, n+l}^{(d_+,d_-)} \\
\kappa_{m,n; m+k, n+l}^{(d_-,p_+)} & \kappa_{m,n; m+k, n+l}^{(d_-,p_-)} & \kappa_{m,n; m+k, n+l}^{(d_-,d_+)} & \kappa_{m,n; m+k, n+l}^{(d_-,d_-)} 
\end{array}\right)\left(\begin{array}{c}
a_{m+k,n+l}^{(p_+)} \\
a_{m+k,n+l}^{(p_-)} \\ 
a_{m+k,n+l}^{(d_+)} \\
a_{m+k,n+l}^{(d_-)}
\end{array}\right)
, 
\end{equation}
where $a_{m, n}^{(o_i)}(z)$ are the amplitudes of the four modes in coupled waveguides, and the elements of the coupling matrices are defined as coupling integrals over the volume of the unit cell: 
\begin{equation}
\kappa_{m,n; m+k, n+l}^{(o_1 o_2)}=\frac{\omega}{c} \frac{1}{\mathcal{P}^{(p)}_{m,n}} \int\left[\varepsilon(\mathbf{r})-\varepsilon_{m+k,n+l}(\mathbf{r})\right] \mathbf{E}_{m,n}^{(o_1) *} \cdot \mathbf{E}_{m+k,n+l}^{(o_2)} d V
,
\end{equation}
where $\omega$ is the frequency of interest, $ \epsilon (r) $ is the permittivity distribution in the entire array consisting of waveguides, $ \epsilon_{m+k,n+l} (r) $ is the permittivity distribution in the waveguide, while $\mathcal{P}_{m,n}^{(p)}=\hat{\mathbf{z}} \cdot \int\left[\mathbf{E}_{m,n}^{(p) *} \times \mathbf{H}_{m,n}^{(p)}+\mathbf{E}_{m,n}^{(p)} \times \mathbf{H}_{m,n}^{(p) *}\right] d V$ is proportional to the power carried by the respective waveguide mode and can be set to unity by the proper normalization of the mode fields; $o_i$ indices enumerate the $p_\pm$ and $d_\pm$ modes of a waveguide ($i = 1...4$).

Although the magnitude of the off-diagonal terms in the coupling matrices $\kappa_{m,n; m+k, n+l}^{(p q)}$ is affected by the relative normalization of the modes, by choosing a specific relative normalization we can ensure that the coupling matrix for each pair of neighboring waveguides is symmetric/antisymmetric~\cite{SavelevGorlach_PRB}.

\begin{figure}[tbp]
    \centering
    \includegraphics[width=0.95\textwidth]{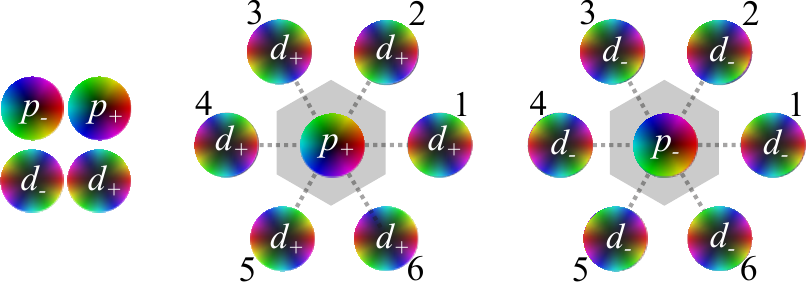}
    \caption{Scheme illustrating the calculation of the coupling integrals for different modes (shown on the left) with two examples of couplings between $p_\pm$ and $d_\pm$ orbital modes. The phase of the field is encoded in color, while radial amplitude profile is arbitrary and serves only for the illustration purposes. 
    }
    \label{FigS1}
\end{figure}

The structure of the coupling matrices is deduced from the mode symmetry using the schematic mode profiles shown in Fig.~\ref{FigS1}. From the definition of the coupling matrices, it follows that $\kappa_{m,n; m+k,n+l}^{(p q)} \propto \textbf{E}_{m,n}^{(p) *} \textbf{E}_{m+k,n+l}^{(q)}$. Considering only nearest-neighbour coupling and azimuthaly invariant intensity of the chosen modes, we recognize that all couplings are governed by the six real-valued parameters $a,b,c,d,e,f$, describing six types of inter-orbital interactions: $p_\pm-p_\pm$, $p_\pm-p_\mp$, $p_\pm-d_\pm$, $p_\pm-d_\mp$, $d_\pm-d_\pm$, $d_\pm-d_\mp$, respectively, while the rest of the degrees of freedom are encoded in phases which could be deduced directly from the phases of pairs of neighbouring modes in the direction of nearest-neighbour coupling, see two examples in Fig.~\ref{FigS1}. In terms of six directions labeled in Fig.~\ref{FigS1} as $1-6$, the corresponding 
coupling matrices read: 
\begin{eqnarray}
&&
\hat{\kappa}_1
=
\left(
\begin{array}{cccc}
 -a & -b & -c & -d \\
 -b & -a & -d & -c \\
 c & d & e & f \\
 d & c & f & e \\
\end{array}
\right),
\hat{\kappa}_2
=
\left(
\begin{array}{cccc}
 -a & b e^{-\frac{i \pi}{3}} & c e^{\frac{2 i \pi }{3}} & d \\
 b e^{\frac{i \pi }{3}} & -a & d & c e^{-\frac{2 i \pi}{3}} \\
 c e^{\frac{i \pi }{3}} & -d & e & f e^{-\frac{2 i \pi}{3}} \\
 -d & c e^{-\frac{i \pi}{3}} & f e^{\frac{2 i \pi }{3}} & e \\
\end{array}
\right),
\hat{\kappa}_3
=
\left(
\begin{array}{cccc}
 -a & b e^{\frac{i \pi }{3}} & c e^{\frac{i \pi }{3}} & -d \\
 b e^{-\frac{i \pi}{3}} & -a & -d & c e^{-\frac{i \pi}{3}} \\
 c e^{\frac{2 i \pi }{3}} & d & e & e^{\frac{2 i \pi }{3}} f \\
 d & c e^{-2 \frac{i \pi}{3}} & e^{-2 \frac{i \pi}{3}} f & e \\
\end{array}
\right), \nonumber \\
&&
\hat{\kappa}_3 = \hat{\kappa}_1^\dag, 
\hat{\kappa}_4 = \hat{\kappa}_2^\dag, 
\hat{\kappa}_5 = \hat{\kappa}_3^\dag
, 
\end{eqnarray}
while a shorter form of them is provided in Eq.~(2) of the main text.

\section{Supplemental Note 2: Accidental chiral symmetry} 

For the special case of equal couplings discussed in the main text ($ a = b = c = d = e = f=\alpha $) our model Hamiltonian features chiral symmetry, manifested in the symmetry of the Bloch spectrum with respect to the zero level [see Fig.~2 of the main text]. The same symmetry holds for the spectra of a finite meta-structure and ribbon [Fig.~3]. 

By definition, chiral symmetry operator $\hat{\Gamma}$ anticommutes with the Hamiltonian. In the original basis of $p_\pm$ and $d_\pm$ modes, it takes the following form 
\begin{eqnarray}
    \hat{\Gamma}
    =
    \left(
    \begin{array}{cccc}
     0 & 0 & 0 & 1 \\
     0 & 0 & 1 & 0 \\
     0 & 1 & 0 & 0 \\
     1 & 0 & 0 & 0 \\
    \end{array}
    \right)
, 
\end{eqnarray}
acting as a permutation operator interchanging $p_+ \longleftrightarrow d_-$ and $p_- \longleftrightarrow d_+$ modes. 

It should be noted that this symmetry is only ``accidental'' in the sense that it is achieved only for such special choice of the couplings, breaking down if the couplings are not equal to each other.

\section{Supplemental Note 3: Symmetry indicators and the topological index}

As discussed in Ref.~\cite{Quantization_2019}, bulk polarization $\textbf{P}$ for $C_6$-symmetric models vanishes. However the topological invariant, $\chi^{(6)} = ([M_1^{(2)}], [K_1^{(3)}])$, may be nontrivial. Here, $[\Pi_{1}^{(2)}] \equiv \# \Pi_{1}^{(2)}-\# \Gamma_{1}^{(2)}$ is an integer topological index where $\# \Pi_{1}^{(2)}$ is the number of energy bands odd under $C_2$ rotation below zero energy at high-symmetry point $\Pi \in \{\Gamma,X,Y,M\}$ in the Brillouin zone. 
The relevant $C_3$ and $C_2$ rotation operators in the basis of $p_\pm$ and $d_\pm$ modes read 
\begin{eqnarray}
    \hat{R}_3
    =
    \left(
    \begin{array}{cccc}
     e^{-2 i \pi/3} & 0 & 0 & 0 \\
     0 & e^{2 i \pi/3} & 0 & 0 \\
     0 & 0 & e^{-4 i \pi/3} & 0 \\
     0 & 0 & 0 & e^{4 i \pi/3} \\
    \end{array}
    \right)
    ,
    \hat{R}_2
    =
    \left(
    \begin{array}{cccc}
     -1 & 0 & 0 & 0 \\
     0 & -1 & 0 & 0 \\
     0 & 0 & 1 & 0 \\
     0 & 0 & 0 & 1 \\
    \end{array}
    \right)
, 
\end{eqnarray}
while both modes and lattice are additionally invariant with respect to $C_6$ symmetry given by operator 
\begin{eqnarray}
\label{R_6}
    \hat{R}_6
    =
    \left(
    \begin{array}{cccc}
     e^{-i \pi/3} & 0 & 0 & 0 \\
     0 & e^{i \pi/3} & 0 & 0 \\
     0 & 0 & e^{-2 i \pi/3} & 0 \\
     0 & 0 & 0 & e^{2 i \pi/3} \\
    \end{array}
    \right)
\end{eqnarray}
in the same basis. 
It is then straightforward to check that $\hat{R}_{2}$ ($\hat{R}_{3}$) commutes with the Hamiltonian Eq.~(1) of the main text at high-symmetry points $\Gamma$, $M$ ($K$), as well as with $C_6$ symmetry operator $R_6$~\eqref{R_6} at $\Gamma$ point. 
The corresponding symmetry eigenvalues are shown in Fig~(2a,c)  near the relevant high-symmetry points for the trivial and nontrivial cases, yielding topogical index $\chi^{(6)} = (2,0)$ in the nontrivial phase.

\section{Supplemental Note 4: Realistic design at microwave frequencies}

\begin{figure}[tbp]
    \centering
    \includegraphics[width=0.75\textwidth]{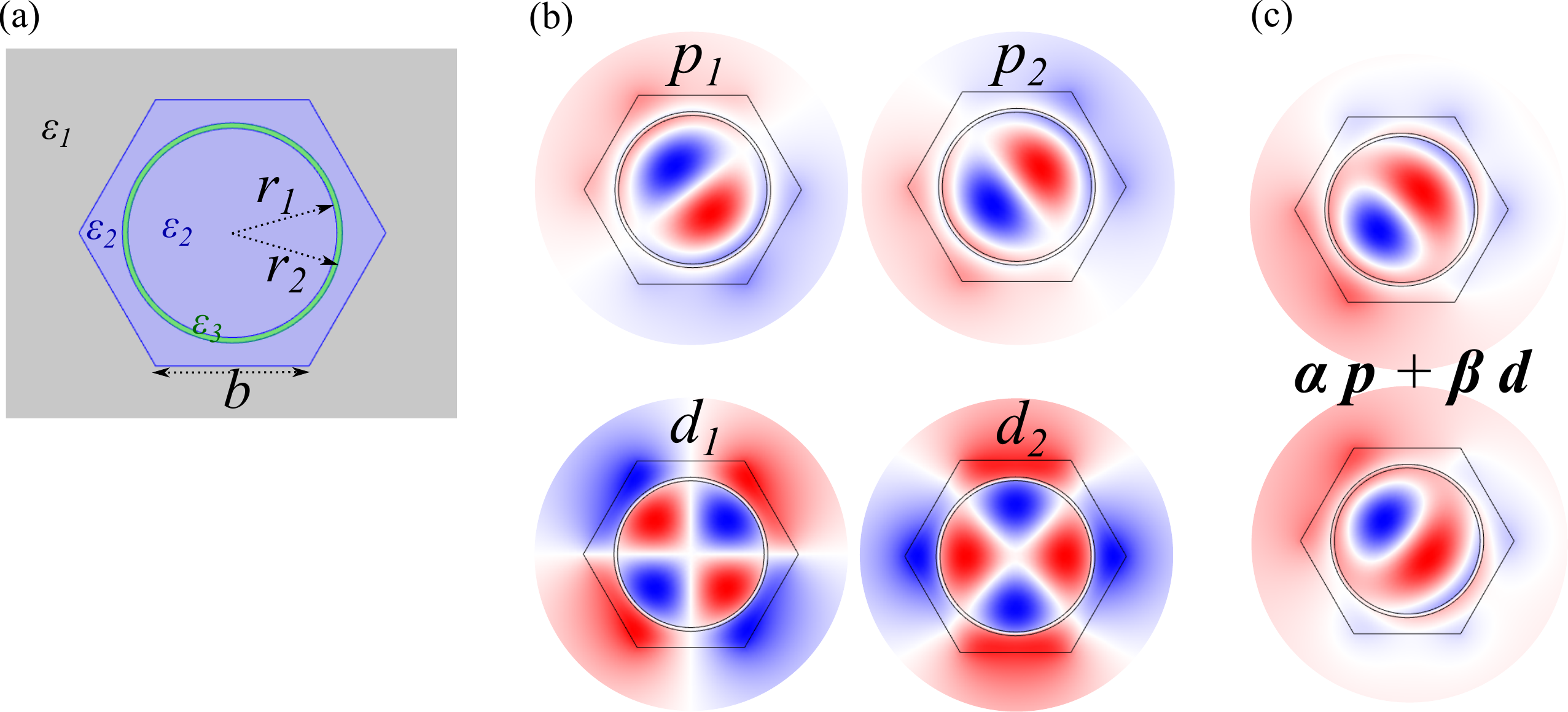}
    \caption{(a) Geometry of a single microwave waveguide with engineered accidental degeneracy between $p$ and $d$ orbital modes in the microwave region at frequency $f = 8.83$~GHz, 
    with $\varepsilon_1=1$, $\varepsilon_2= 8.0 $ and a metallic inclusion with $\varepsilon_3=0.2$. 
    Common parameters: $b=0.014$~m, $r_1 = 0.954$~cm  and $r_2 = 1.002$~cm. 
    (b) Field profiles of the degenerate modes 
    (shown here is the dominant component of the electric field $E_z$ responsible for the inter-orbital couplings). 
    (c) Two linear equally-weighted superpositions ($|\boldsymbol{\alpha}| \sim |\boldsymbol{\beta}|$) of the four degenerate $p_\pm$ and $d_\pm$ modes, resembling of corner mode profiles in Fig.~(3b) in the main text. 
    }
    \label{FigS2}
\end{figure} 

\begin{figure}[tbp]
    \centering
    \includegraphics[width=0.95\textwidth]{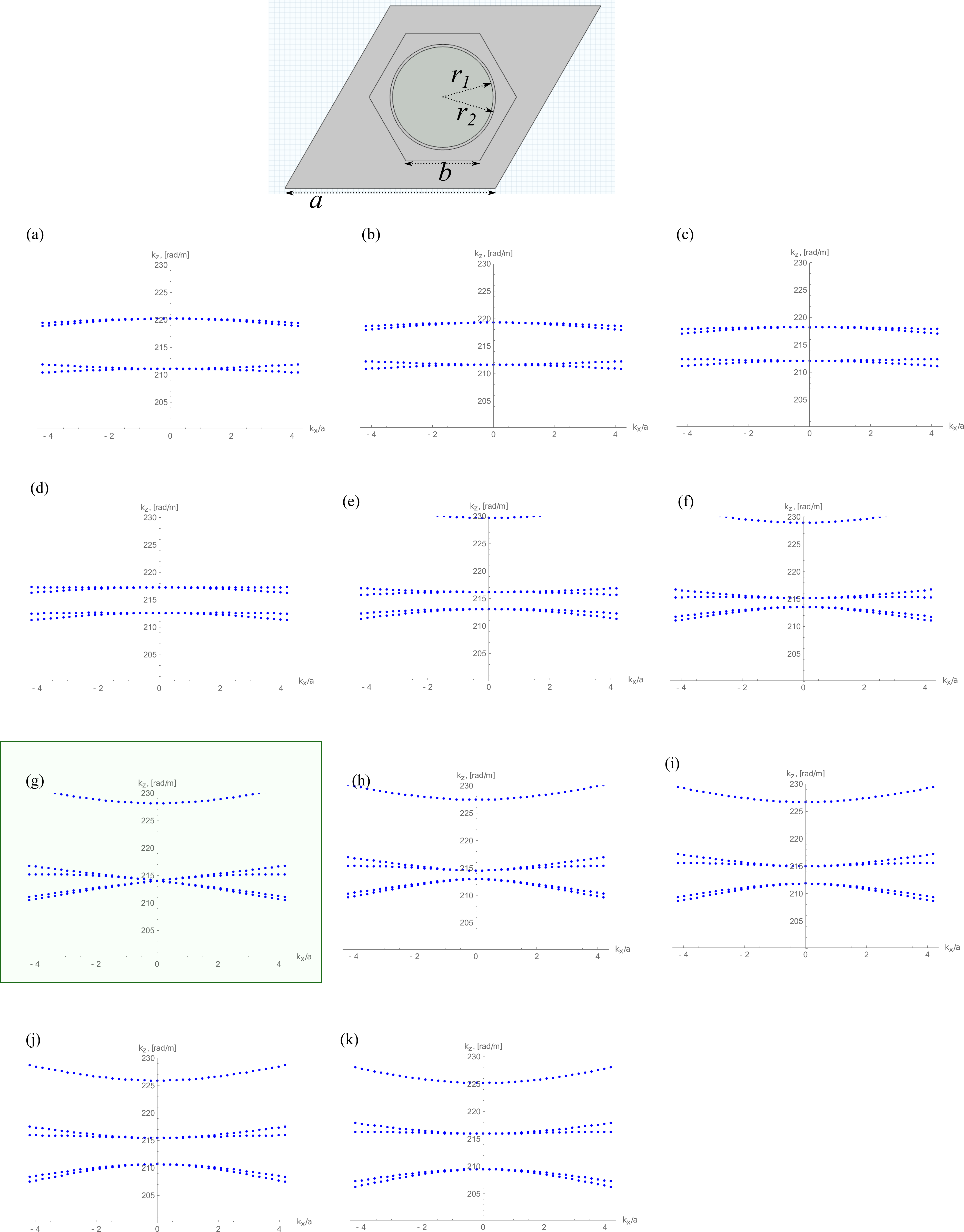}
    \caption{Mode dispersion calculated numerically with periodic boundary conditions (see geometry in the top inset) near the $\Gamma$ point (varying $k_x$ and fixed $k_y=0$) for different geometries of the waveguides. Common parameters: $a=0.04$~m, $b=0.014$~m, $r_1 = 0.95425 - 0.00887 \psi$~cm  and $r_2 = 1.00225 - 0.00887 \psi$~cm, with sweep in $\psi$ from $-5$ to $0$. 
    Panel (g) shows gapless double-Dirac-cone spectrum for $\psi = -2$, in full correspondence with theoretical predictions for the triangular $p$-$d$-hybridized lattice~\cite{Sakoda_2012_1,Sakoda_2012_2}. 
    }
    \label{FigS3}
\end{figure} 

To demonstrate the feasibility of our proposal, we consider a microwave waveguide with hexagonal cross-section having a circular cutout (see Fig.~\ref{FigS2}a). Such geometry of the waveguide ensures the degeneracy of $p$-like (HE$_{12}$) and $d$-like (HE$_{22}$) modes at frequency $f = 8.83$~GHz 
for a fixed geometry, shown in Fig.~\ref{FigS2}b, with propagation constants around $214.5 \,\text{rad/m}$ and detuning between the propagation constants $\delta\beta$ of the order of $0.5 \,\text{rad/m}$. 
Note that although this particular geometry features hexagonal waveguides and the metallic inclusion with $\varepsilon_3=0.2$, similar degeneracy could be achieved also in circular waveguides with a simple air gap with $\varepsilon_3=1$. The parameters chosen here are tuned additionally \textit{for the clarity of demonstration} to ensure: (1) maximal photonic topological bandgap; (2) consistently large spectral distance of the four modes of interest $p_\pm$ and $d_\pm$ from other-order modes for changing geometric parameters (see the geometric tuning parameter $\psi$ below and in Fig.~\ref{FigS3}). 
The dominant electric field component responsible for the mode coupling in this case is the longitudinal component of electric field, $E_z$. 
To implement the physics discussed in the main text it is also  possible to utilize waveguides with a cutout and a symmetry higher that $C_6$, e.g. $C_{12}$-symmetric or circular-shaped waveguides. 
Permittivities and geometric parameter $\psi$ were chosen in such a way that the the chosen four $p_\pm, d_
\pm$ modes are always spectrally far away from other waveguide modes so that the behavior of the lattice is properly captured by the four-mode tight-binding physics.


Next, we calculate numerically and examine qualitatively the band diagram of the two-dimensional double-periodic array. 
The band diagram obtained from full-wave calculations with periodic boundary conditions is shown in Fig.~\ref{FigS3} (with sweep in $k_x$ while keeping $k_y=0$). 
We observe that for a certain fine-tuned cutout geometry (see Fig.~\ref{FigS3}), the band diagram features a double-Dirac-cone spectrum, as expected from our tight-binding theory (see Fig.~(2b) in the main text) as well as previous group-theory theoretical predictions~\cite{Sakoda_2012_1,Sakoda_2012_2}. 
This additionally confirms the validity of our simplified four-band tight-binding model. 

Finally, we calculate the spectrum for a a finite hexagonal lattice, as shown in Fig.~(3b) in the main text. For the geometric parameters in the topological regime ($\psi=0$), the spectrum hosts six hybrid-orbital corner states in the bandgap as well as edge states around the bandgap, in full agreement with the tight-binding analysis.



\bibliography{refs}